\documentclass[useAMS,usenatbib]{mn2e}
\usepackage{times}
\usepackage{color}
\usepackage{graphicx}
\title[Morpho-kinematics of NGC\,3242]
{Morpho-kinematics of the planetary nebula NGC\,3242: an analysis beyond its multiple-shell structure\thanks{Based upon observations acquired at the Observatorio Astron\'omico Nacional in the Sierra San Pedro M\'artir (OAN-SPM), Baja California, Mexico.}}
\author[M. A. G\'omez-Mu\~noz et al.]
{
M. A. G\'omez-Mu\~noz$^{1}$\thanks{E-mail:marcog@astro.unam.mx},
M. W. Blanco C\'ardenas$^{1}$,
R. V\'{a}zquez$^{1}$, 
S. Zavala$^{2,3}$, 
\newauthor
P. F. Guill\'en$^{1}$ and S. Ayala$^{4}$\\
$^{1}$Instituto de Astronom\'{\i}a, Universidad Nacional Aut\'onoma de
M\'exico, Apdo. Postal 877, 22800 Ensenada, B. C., Mexico\\
$^{2}$Instituto Tecnol\'ogico de Ensenada, Blvd. Tecnol\'ogico No. 150, 22780 Ensenada, B. C., Mexico\\
$^{3}$Instituto de Estudios Avanzados de Baja California, A. C., Blvd. Tte. Azueta 147, Edif. Matematik\`e Planta Baja, 22800 Ensenada, B. C., Mexico\\
$^{4}$Facultad de Ciencias F\'{\i}sico-Matem\'aticas, Universidad Aut\'onoma de Nuevo Le\'on, Av. Universidad s/n, 66451 San Nicol\'as de los Garza, N. L., Mexico\\
}

\begin{document}

\date{Accepted 2015 August 18. Received 2015 August 17; in original form 2015 May 4}

\pagerange{\pageref{firstpage}--\pageref{lastpage}} \pubyear{}

\maketitle

\label{firstpage}
\begin{abstract} 

In this paper we present the results of optical high-resolution imaging and spectroscopy of the
complex planetary
nebula (PN) NGC\,3242. Our study is based on the analysis of
the narrowband H$\alpha$ $\lambda6563$\AA,
[O\,{\sc iii}] $\lambda5007$\AA, [N\,{\sc ii}] $\lambda6584$\AA, and [S\,{\sc ii}] $\lambda6724${\AA} images, and high-resolution
spectroscopy using spectral ranges centered on the H$\alpha$ $\lambda6564$\AA, [N\,{\sc
  ii}] $\lambda6583$\AA, and
[O\,{\sc iii}] $\lambda5007${\AA}. We detected and analysed
morphological components beyond the multiple shell structure of this
PN, to investigate the
small-scale morphological components aligned towards its major axis (such as
knots and ansae, as well as the arc-like features) 
and its surroundings. Thus, we investigated the morpho-kinematical properties of NGC\,3242, as well as
their nature and formation.
Our results regarding the elliptical double-shell structure and the
distance to this nebula are in concordance with previous studies.
Furthermore, we have used the software {\sc shape} to construct a 3D model of
NGC\,3242, allowing us to successfully reproduce our observational
data. 
We conclude that the prominent knots
emitting in the [N\,{\sc ii}] line are
fast, low-ionisation emission regions (FLIERs) related to high velocity jets and the
so-called ansae-like features rather resemble bubbles. The
disruptions immersed in the halo, whose emission was detected in the 
the [O\,{\sc iii}] high-excitation emission line, remarkably display high velocities and
were formed likely in an earlier ejection event, in comparison to the
innermost FLIERs and bubbles. Finally, according to our model, the kinematical ages of the structures in NGC\,3242
range from 390 to 5400\,yr.
\end{abstract}


\section{Introduction}

Planetary Nebulae (PNe) evolve from low and intermediate-mass stars (0.8 -- 8 M$_\odot$)
and mainly consist of an ionised gaseous envelope surrounding a stellar
nucleus. The shaping of a PN takes place at some point between the asymptotic giant branch 
(AGB) and the white
dwarf stellar phases. The characteristic ionised envelope of a PN results from the strong mass-loss
experienced during the AGB and post-AGB phases through the stellar
wind ejected by their progenitor stars. The interacting stellar wind
model \citep[ISW,][]{kwo78} and its generalisation
\citep[GISW,][]{bal87} provide a good explanation to the shaping of the spherical,
elliptical, and even slightly bipolar morphologies. However, the
formation of more complex morphologies often displaying collimated outflows, high velocity jets,
and point symmetric features remains under debate 
\citep[see][for a review]{ba2002}.

Beyond the ionised envelope and the morphological structures therein,
huge haloes have been found in several evolved PNe, giving them a
triple-shelled structure. These haloes are ionised by D-type fronts and are the
remnant of the AGB envelope \citep{sch03}. Furthermore, imaging using the Hubble Space Telescope (\textit{HST}) have revealed the
presence of features resembling rings and arcs immersed in the
haloes of four PNe: Hb\,5, NGC\,6543, NGC\,7027 \citep{ter97} and
NGC\,3918 \citep{cor03b}. As for ground-based imaging,
\citet{cor04} have exposed the presence of these arc-like
features in NGC\,40 and NGC\,3242 using the 2.5-m Isaac Newton
Telescope (INT). These rings and arc-like structures in the haloes likely arise at the end of the AGB phase,
during the last 10,000 to 20,000 years, however, according to
\citet{cor04}, this phenomenon is not related
with the thermal pulses.

NGC\,3242 (a.k.a. `The Ghost of Jupiter') is a multiple-shell PN.
The brightest inner ellipse (hereafter referred as inner shell)
has a size of  28$\times$20\,arcsec and, according to
\citet{bal87}, is expanding at $\simeq$30\,km\,s$^{-1}$ and has a
density of 2,200\,cm$^{-3}$ \citep{nie06}. This inner shell is surrounded by an expanding, fainter, moderately
elliptical shell (hereafter referred to outer shell) of 46$\times$40\,arcsec in size whose density declines 
$\propto r^{-1/2}$ \citep{nie11}. 
Between the inner shell and the outer shell of NGC\,3242, the presence of
a bright pair of knots detected in [N~{\sc ii}] and two ansae-like structures
orientated towards NW-SE is remarkable \citep{bal98}. 
The observations of \citet{nie11} using
XMM-Newton have shown the X-ray luminosity is $\simeq 2 \times
10^{30}$ erg s$^{-1}$ for a distance of 0.5 kpc
\citep{ter97}. The electronic temperature of the hot bubble
region, in which the X-rays are produced, has a value of 
$\simeq 2.35\times10^{6}$\,K.

These two shells are enclosed by a broken halo revealed by deep images
\citep{cor03b,cor04} and at least three rings and two arc-like features are observed
within this halo. \citet{mon05} have found an apparent gradient in the electronic
temperature of the halo increasing towards its outer regions
(15,700\,K$< T_{e} <$20,300\,K). \cite{bla06} and  \cite{phi09} also noted some arc-like features in the halo, disrupting the rings. 
Given the location and orientation of these features, they are
probably associated with the bright [N~{\sc ii}] knots and ansae-like
structures observed between the inner shell and the outer shell.


NGC\,3242 also shows an outermost huge halo
detected by means of \textit{Spitzer} imaging
\citep{ram09}. The characteristics of this asymmetric
halo, composed by gas and thermal dust grains, extending towards the north,
suggest an interaction of NGC\,3242 with the ISM.
Despite the kinematics of NGC\,3242 has been studied before \citep{bal87, mea00, ro10}, in this
paper we pursue a detailed analysis of the small-scaled structures
detected along the major axis, distributed in the multiple-shell
structure of this PN, focusing our
attention in the 
puzzling
arc-like features in the halo.

\section{Observations}

\subsection{CCD direct images}
\label{imagenes}


Narrow-band CCD direct images
of NGC\,3242 were obtained on 2004 May 9 and 10 using the 1.5-m ($f/13$)
Harold Johnson Telescope at the San Pedro M\'artir Observatory (OAN-SPM\footnote{The Observatorio 
Astron\'omico Nacional (OAN-SPM) is located at the Sierra de San Pedro M\'artir, Baja California, 
and is operated by the Instituto de Astronom{\'\i}a of the Universidad Nacional Aut\'onoma de M\'exico 
(UNAM).}). The detector was a 1024$\times$1024 SITe CCD (24\,$\mu$m pixel size) with a plate 
scale of 0.25\,arcsec\,pix$^{-1}$. 
The filters used to acquire the images were H$\alpha$ ($\lambda_c = 6563$\AA, $\Delta \lambda$ $= 10$\AA ),
[O\,{\sc iii}] ($\lambda_c = 5007$\AA, $\Delta \lambda$ $= 50$\AA),
[N\,{\sc ii}] ($\lambda_c = 6584$\AA, $\Delta \lambda$ $= 11$\AA ), and
[S\,{\sc ii}] ($\lambda_c = 6724$\AA, $\Delta \lambda$ $= 10$\AA). 
Exposure times were 180\,s for [O\,{\sc iii}] and
600\,s for the 
other filters. Seeing was around 2-arcsec during observations.
Images were processed using standard 
techniques of the image reduction and analysis facility package ({\sc iraf}).
Figure \ref{mosaico} shows  all the images. Labels are explained in Sect. \ref{results}.

\subsection{Long-slit echelle spectroscopy}

High-resolution, long-slit spectra of NGC\,3242 were obtained on different observing runs. Observations were carried out with the Manchester Echelle Spectrograph (MES; \citealt{mea03}) in the 2.1-m ($f/7.5$) telescope at OAN-SPM. A e2v 13.5-$\mu$m\,pix$^{-1}$ CCD with $2048\times2048$ pixels was used as detector. The slit width was 150\,$\mu$m (1.9\,arcsec). Slits positions are shown in Figure 2 and are labeled as s1 to s12. Position angles (PAs) for these observations were +25{\degr}
(s2, s12), +60{\degr} (s1, s11), +56{\degr} (s3--s9), and $-30\degr$
(s10).

The first observation run was executed on 2013 December 15 using the $4\times4$ binning mode (0.702 arcsec\,pix$^{-1}$ plate scale). The spectra were centered at 
the [O\,{\sc iii}]$\lambda5007$ emission line using a filter ($\Delta\lambda$=50\AA) to isolate the 114$^{\rm th}$ order (0.087 {\AA}\,pix$^{-1}$  spectral scale). Exposure time was 1800\,s for each spectrum (slits s2 and s12). 

The second series of spectra were acquired on 2014 January 18 in the
$2\times2$ binning mode (0.351 arcsec\,pix$^{-1}$ plate scale). Observations were made around
the H$\alpha$ emission line using a filter ($\Delta\lambda$=90\AA) to isolate the 87$^{\rm th}$ 
order (0.1 {\AA}\,pix$^{-1}$  spectral scale). Exposure time was set to 1800\,s. Three slit positions
 were located to pass through the bright [N\,{\sc ii}]$\lambda$6584 knots as well as the 
arc-like features avoiding the central star to prevent saturation of the spectra (s1, s10, and s11).
Furthermore, slit s10 pass through the major axis and was observed in H$\alpha$ and [O\,{\sc iii}]$\lambda5007$ spectral ranges, but with a spectral 
scale of 0.043 {\AA}\,pix$^{-1}$ and exposure time of 1200\,s.

Finally, the third series of spectra was  observed on 2014 January 20 with the
same technical configuration as the December run. Exposure times were 300\,s with the exception 
of s9 that was 600\,s. Seeing was around 1.5~arcsec during the observations  (s3 to s9).

Data were processed using standard techniques for long-slit spectroscopy of {\sc iraf}. 
The spectra were wavelength-calibrated with a Th-Ar arc lamp to an accuracy of $\pm1\,{\rm km\,s}^{-1}$.
The FWHM of the arc lamp emission lines was measured to be $\simeq12\,\pm1\,{\rm km\,s}^{-1}$. 

\section{Results}
\label{results}

In the following two subsections we present our main results separated in the inner and outer structures, with regard
to the two shells and the halo. We also estimated the distance to NGC\,3242.

\subsection{Inner structures}

\label{mor}

\begin{figure*}
\centering
\includegraphics[width=0.625\textwidth]{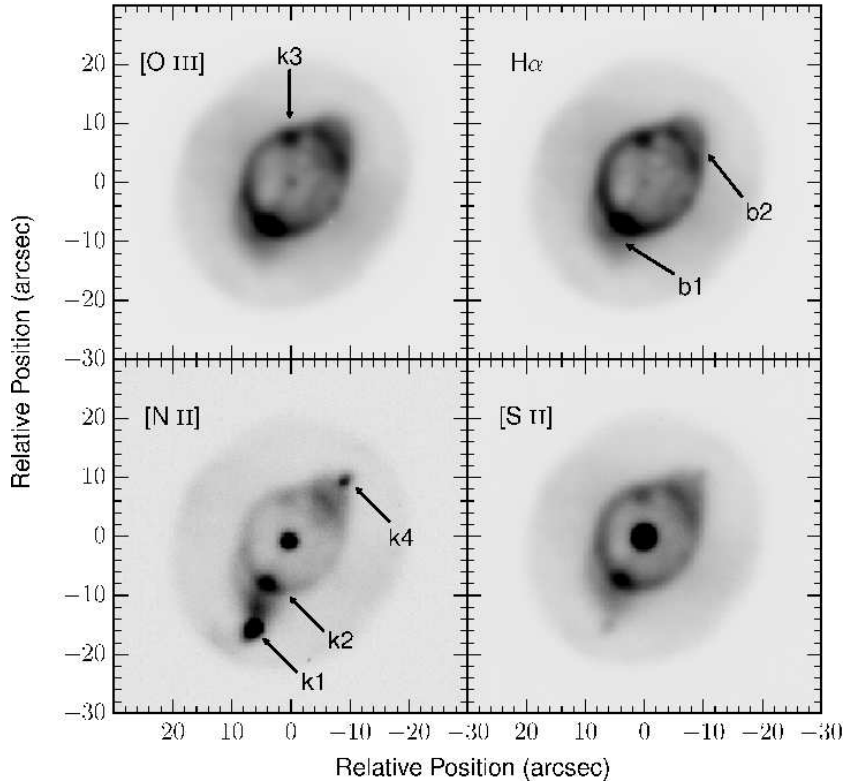}
\caption{NGC\,3242 as seen in (a) H$\alpha \lambda$6563{\AA}, (b) [O\,{\sc iii}]$\lambda$5007{\AA}, (c) [N\,{\sc ii}]$\lambda$6584{\AA}, and (d) [S\,{\sc ii}]$\lambda$6724{\AA}. Arbitrary contrast scales were set to enhance the different features seen in the nebula. North is up and East is left in all the images.
\label{mosaico}}
\end{figure*}

\begin{figure}
\centering
\includegraphics[width=0.47\textwidth]{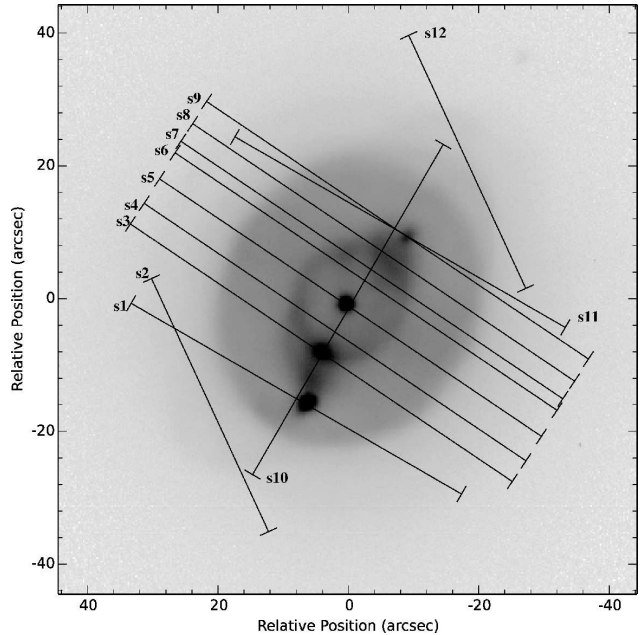}
\caption{Slits position observed and superimposed in the [O\,{\sc iii}] + [N\,{\sc ii}] composite image with arbitrary contrast chosen to enhance the different features present in the nebula. PAs=+25\degr (s2, s12), +60\degr (s1,s11), +56\degr (s3--s9), and --30{\degr} (s10).}
\label{slits}
\end{figure}

As can be seen in our direct imaging study, some of the structural
components of NGC\,3242 appear slightly different according to the
wavelength observed (Fig. \ref{mosaico}). We detected the two
concentric elliptical components: the limb brightening inner shell
 (28$\times$20\,arcsec)
surrounded by a diffuse outer shell 
(46$\times$40\,arcsec), structures
previously reported by
\cite{hro68},
in which 
the major axis are located toward the PA=$-40$\degr.

Besides the inner and outer shells, further small structural components were
resolved and, as mentioned before, were detected in different wavelengths.
 These structures are labeled in Fig. \ref{mosaico}.

\begin{figure*}
	\centering
  \includegraphics[width=0.7\textwidth]{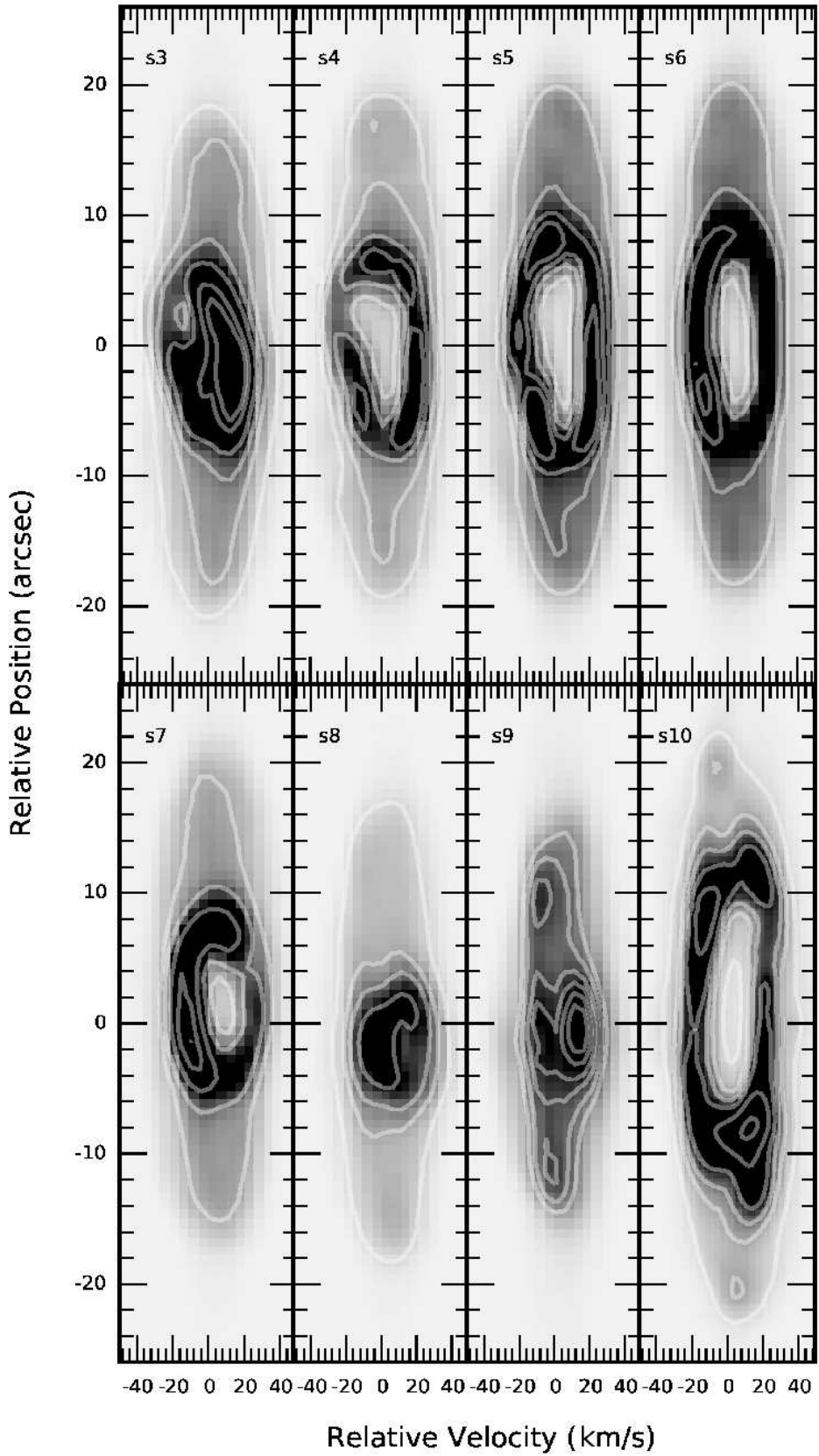}
  \caption{Position-Velocity (PV) maps of the [O\,{\sc iii}] emission line corresponding to 
 the s3 -- s6 (upper panel) and s7 -- s10 (lower panel). NE is up in all the panels, except s10, in which
 NW is up.
 \label{specO3}}
 \end{figure*}

\begin{figure}
 \centering
 \includegraphics[width=0.47\textwidth]{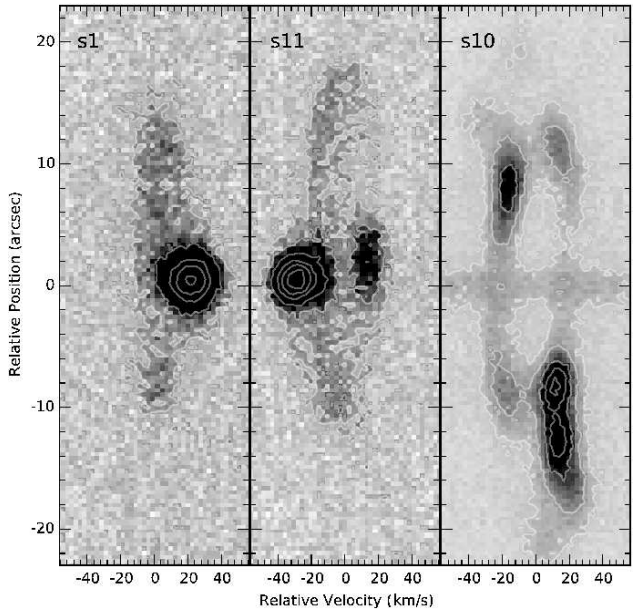}
 \caption{Position-Velocity maps in [N\,{\sc ii}]. 
 Labels correspond to slit position shown in Fig.~\ref{slits}: s1 (left), s11 (middle) and s10 (right). Position axis is reference to the peak of emission (s1 and s11) and to the central star (s10). NE is up in the panels left and center, whereas NW is up in the right panel.
 \label{specHa}}
\end{figure}

 About kinematics, Figure \ref{specO3} shows Position-Velocity (PV) maps corresponding 
to the slits s3 to s10, whereas those from s1, s11, and s10 are presented in Figure \ref{specHa}.  
The radial velocity presented hereby is relative to the systemic velocity 
$V_{\rm LSR} = -6.6\pm 1$\,km\,s$^{-1}$ ($V_{\rm Hel} = +1 \pm
1$\,km\,s$^{-1}$) obtained from s10 and which is in good agreement with 
the value of $\simeq -7.6 \pm  1$\,km\,s$^{-1}$ obtained by \citet{ro10}.
A well-defined velocity
ellipsoid dominates the velocity pattern measured, however,
distortions corresponding to the inner structures are present and
enhanced in our PV diagrams by contours. 
The measured expansion
velocity of the inner shell is 23 $\pm 1$\,km\,s$^{-1}$ and is
consistent with the value obtained by \cite{bal87}.  As for the outer
shell, we have 
estimated 
an expansion velocity of 
$\simeq13 \pm 1$\,km\,s$^{-1}$  by considering the deprojected separation of two Gaussian components
fitted to a line profile from s6 (less deformed velocity ellipsoid), taken at 14 arcsec from the central star in order
to avoid confusion from the inner shell. Our value is a typical for this type of outer shell.


Table \ref{tabla1} resumes the main parameters of structures labeled in Fig. \ref{mosaico}.
Column 1 shows the label of each microstructure (k, for knot, b, for blob). Columns 2 and 3 correspond
to the slit and filter used to measure the velocity. Column 4 and 5 indicate the angular distance
from the central star, as well as the PA, measured from north to east, for each structure. The radial
velocity relative to the systemic velocity is presented in column 6. In the case of the b1 and b2, which
also present an expansion, the values in the table represent position and velocity of the center of each
blob. unnecessary 

In general our results are consistent with previous works in literature.

\begin{table}
	\centering
	\begin{tabular}{|clc|c|c|c|c|}
		\hline
		Name	&	Slit		&	Filter		&	Distance	&	PA		&	$V_{\rm rel}$	\\ 
				& 			& 				&	(arcsec)	&	(\degr)	&	(km\,s$^{-1}$)				\\ \hline
		
		k1	&	s1	&	[N\,{\sc ii}]		&	16		&	$+$158	&	+22$\pm$1	\\
		k2	&s3,s10& [O\,{\sc iii}], [N\,{\sc ii}]	&9.5&	$+$152	&	+14 		\\
		k3	&	s7	&	[O\,{\sc iii}] 	&	7.5		&	$+$4		&	+2.3		\\
		k4a&	s11&	[N\,{\sc ii}]	&	14			&	$-$40		&	$-$28	\\
		k4b&	s11&	[N\,{\sc ii}]	&	14			&	$-$40		&	+13		\\
		b1	&	s3	&	[O\,{\sc iii}]	&	8.7		&	$+$140	&	$-$13	\\
		b2	&	s8	&	[O\,{\sc iii}]	&	8.7		&	$-$40		&	+11		\\		

		\hline
	\end{tabular}
	

	\caption{Small structures properties of NGC\,3242. \label{tabla1}}
\end{table}

From Fig. \ref{specO3} we have also been able to
measure the expansion velocity of {\bf b1} and {\bf b2}  (see slits s3 and s8, respectively) and the values obtained were
21$\pm 1$\,km\,s$^{-1}$ and 16$\pm 1$\,km\,s$^{-1}$, respectively. These values were calculated by fitting a double-peaked Gaussian profile
to the spectral lines corresponding to each bubble and considering the contamination of the emission of the inner shell.

\subsection{Outer structures}
\label{kinematics}


\begin{figure}
 \centering
 \includegraphics[width=0.47\textwidth]{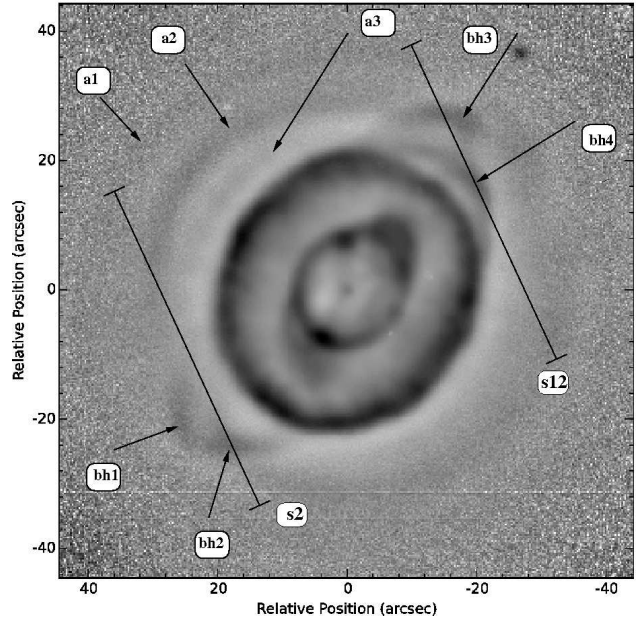}
 \caption{Unsharp-masking logarithmic image in [O\,{\sc iii}] of NGC\,3242. 
Main morphological features are labeled.}
 \label{umask2}
\end{figure}

 \begin{figure}
	\centering
  \includegraphics[width=0.30\textwidth]{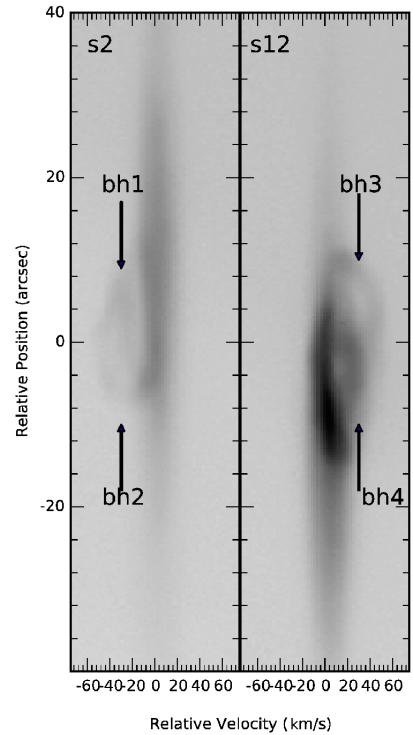}
  \caption{Position-Velocity (PV) maps of the [O\,{\sc iii}] emission line corresponding to 
 the s2 (left), and s12 (right) slit positions. NE is up in both panels.
 \label{specO3b}}
 \end{figure}

In order to enhance the outermost morphological components of NGC\,3242,
those located in the halo, we applied the widely used unsharp masking
technique to the [O\,{\sc iii}] image (Fig. \ref{umask2}). Despite the
inner and outer shells, as well as the features {\bf k2}, {\bf k3}, {\bf b1},
{\bf b2}, dominate the emission of this PN, at least three arcs ({\bf a1}, {\bf a2} and
{\bf a3}) are detected beyond the concentric shells.
Furthermore, it is remarkable the presence of two
pairs of features aligned once more towards NW--SE:
{\bf bh1}, {\bf bh2}, {\bf bh3}, {\bf bh4}. These features were previously detected by \cite{bla06} and \cite{phi09}.
Henceforward, we will refer to these four features as ``disruptions'' of the halo.
 The kinematics of the disruptions were investigated in the PV diagrams of the
Figure \ref{specO3b}, that is, in the slits s2 and s12.  These structures appear like
loops in the PV maps. Radial velocity of the central point of each structure, relative
to the systemic velocity of the nebula, are shown in Table \ref{tabla2}, as well as the
expansion velocity ($V_{\rm exp}$) on each case. For each structure, radial velocities are estimated
from the center of the loop (mean velocity of the splitting), whereas expansion
velocities are estimated as the half of the maximum of each loop (line splitting) in
the velocity axis.

\begin{table}
	\centering
	\begin{tabular}{|clc|c|}
		\hline
		
		Structure	&	$V_{\rm rel}$	&	$V_{\rm exp}$	\\
					&		(km\,s$^{-1}$)			&	(km\,s$^{-1}$)		\\ \hline
					
		bh1		&		$-$19$\pm$1			&	11$\pm$1		\\
		bh2		&		$-$29					&	5				\\
		bh3		&		+29						&	5				\\
		bh4		&		+13						&	12				\\

		\hline
	\end{tabular}
	

	\caption{Outer structures properties of NGC\,3242. \label{tabla2}}
\end{table}



\subsection{Distance to NGC\,3242}
\label{distance-parallax}

Trigonometric parallax for NGC\,3242 was early determined by Jenkins (1952), who estimated a value of $30\pm15$~mas. Such value lead to an unreal determination of distance for this nebula. Subsequently, \cite{cah92} reported a distance of 1.08 kpc by means of statistical methods. Using VLA data and an expansion parallax method, \cite{haj95} determined a distance to NGC\,3242 of  $420\pm160$\,pc. In this case, data time lapse was around 6 years. A correction to the parallax method was calculated by \cite{mel04}, lead to a corrected distance of $550\pm230$\,pc to NGC\,3242.

In this work, we have estimated a new value for the distance to NGC\,3242 using HST imaging from two different epochs, as well as the equatorial expansion velocity described in the Section \ref{kinematics} and the {\sc shape} model presented in Section \ref{model-shape}.

\begin{figure}
\centering
\includegraphics[width=0.35\textwidth]{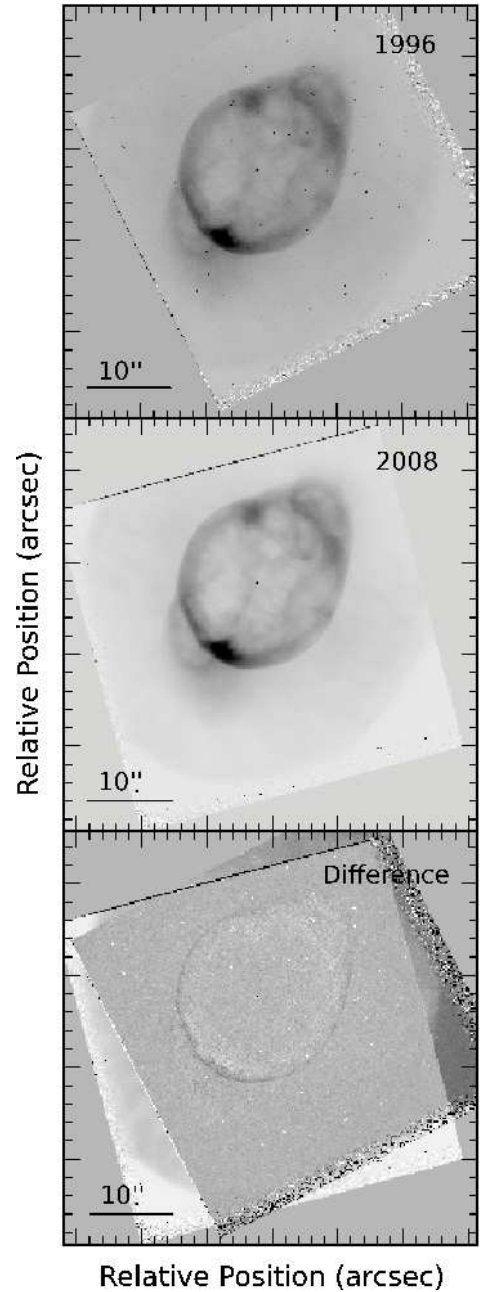}
\caption{[O\,{\sc iii}] (F502N) HST images from the MAST archives: programs 6117 [1996] (Top) and 11122 [2008] (middle). The bottom image is the difference of the two HST images.}
\label{dist-calc}
\end{figure}

We got the [O{\sc iii}] (F502N, $\lambda_{\rm c}=5012${\AA}, $\Delta \lambda=27${\AA}) HST images from the MAST archives (programs 6117 [1996] and 11122 [2008], P. I. Bruce Balick). Measurements of the equatorial size along the minor axis of each image (PA=+50\degr) were performed with the task {\it splot} of {\sc iraf}, co-adding 10 rows to enhance  the signal-to-noise ratio. Borders of both sides were de-blended and the extreme components were used to determine the diameters on each image. A difference of 3.44 pixels in diameter (1.72 in radius) was measured. The two different epochs images, as well as the image of the difference, supporting the real nature of the expansion, are presented in Figure \ref{dist-calc}.

We have determined the distance to NGC\,3242 considering a constant velocity and using the following equation, which includes physical units transformation.

\[
\left[ {r \over {\rm pc}} \right] = 4.22 \left(
{ \left[ {v_{\rm exp}\over {\rm km\,s^{-1}}}\right] \,
   \left[ {t\over {\rm yr}} \right] \over \left[ {s\over {\rm pixels}}\right] }
   \right)
\]

Using the expansion velocity of 23$\pm$1\,km\,s$^{-1}$, the time lapsed as 11.7$\pm$0.04\,yrs and the 1.72$\pm$0.5\,pixels from the last paragraph, we have estimate a distance of $660\pm100$\,pc, which is in agreement with the determinations of \cite{haj95} and
even better when comparing with the corrected value from
 \cite{mel04}. It is clear that the uncertainty of the radius dominates the error of this calculation.

\section{Discussion}  
\label{discusion}

\subsection{The morpho-kinematic structure of NGC\,3242}
\label{model-shape}

\begin{figure*}
 \centering
 \includegraphics[width=0.8\textwidth]{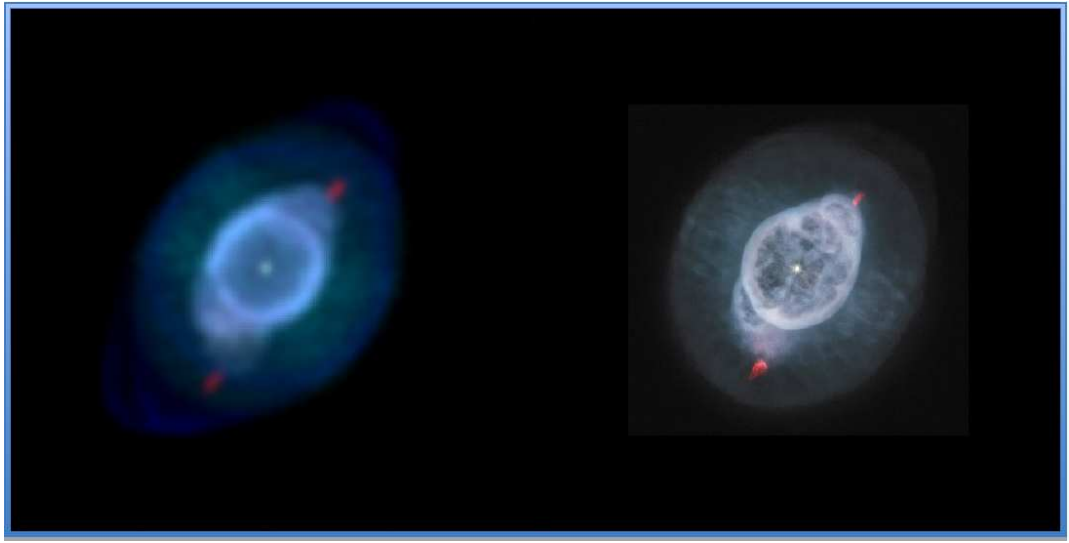}
\caption{Comparison of the 3D model obtained from {\sc Shape} (left) and the HST image (right).}
\label{mosaico-shape}
\end{figure*}

As described above (Sec.~\ref{mor}), our analysis of the high-resolution images indicates that NGC\,3242 is a 
morphologically rich PN with ansae-like features, that resemble bubbles, and knots inside its main structure (Fig.~\ref{mosaico}). In a first 
attempt to study the morphology and kinematics of NGC\,3242, we have constructed a geometrical model using 
the interactive software {\sc shape} \citep{ste11}, a computational tool for morpho-kinematic modeling and 
reconstruction of astrophysical objects. The internal library of {\sc shape} includes basic structures such as 
spheres, cylinders, and torii, which can be modified by applying different parameters and/or functions (`modifiers'). 
Physical parameters, such as density and velocity, can be assigned to the structures either as an analytical 
function or interactively. Several recent papers have used {\sc shape} to get the structure of PNe 
\cite[e.g.,][]{mis11, ram12, vaz12, dan15}. In our particular case, we obtain a final nebula model that fits 
the main structure of NGC\,3242 for both imaging and spectroscopy (Fig. \ref{mosaico-shape}).

\begin{figure}
 \centering
 \includegraphics[width=0.45\textwidth]{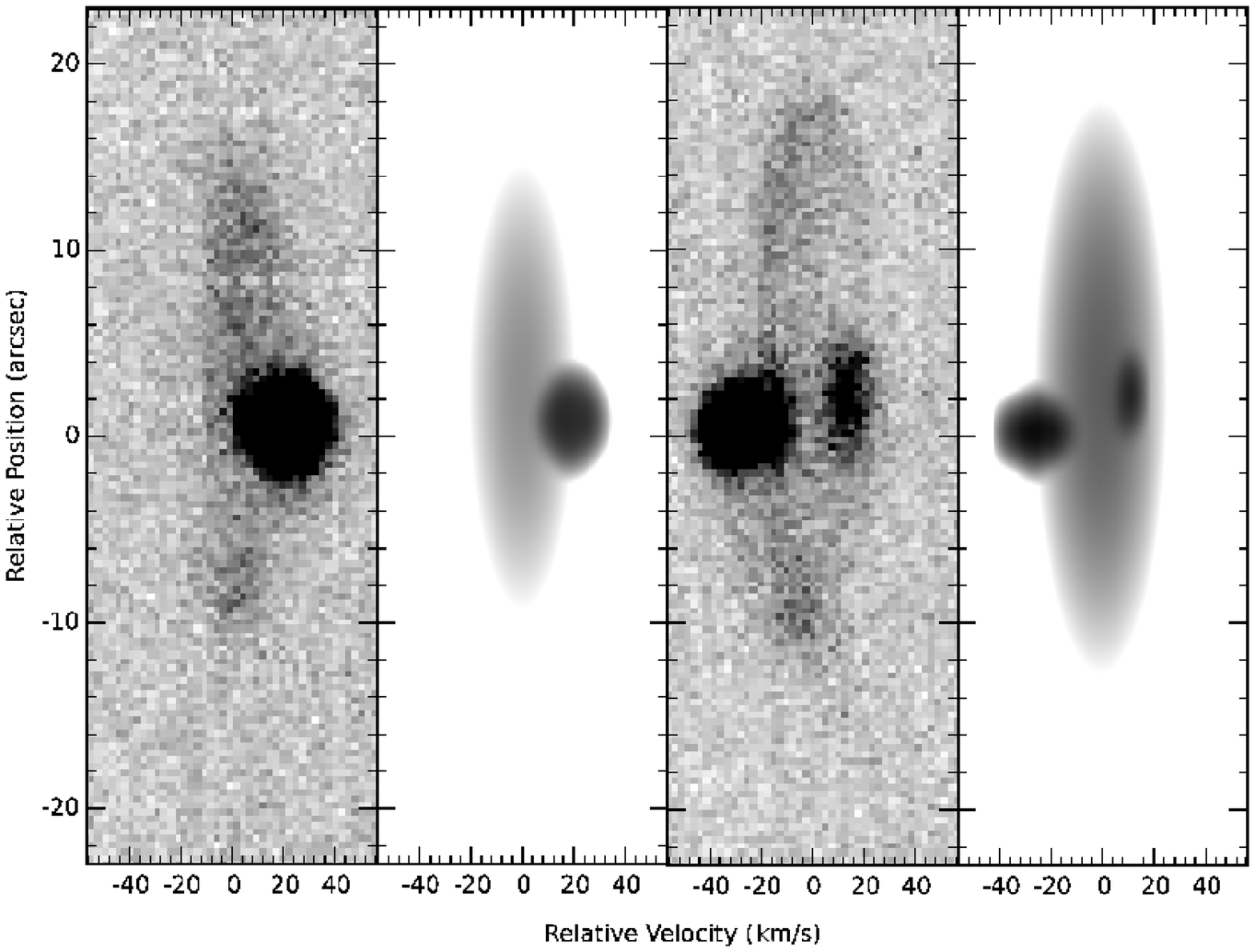}
\caption{Observed PV-map of the [N\,{\sc ii}] line for s1 and s11, and the corresponding {\sc shape} synthetic 
PV-maps. NE is up in all the panels.}
\label{jets-shape}
\end{figure}

\begin{figure}
	\centering
	\includegraphics[width=0.45\textwidth]{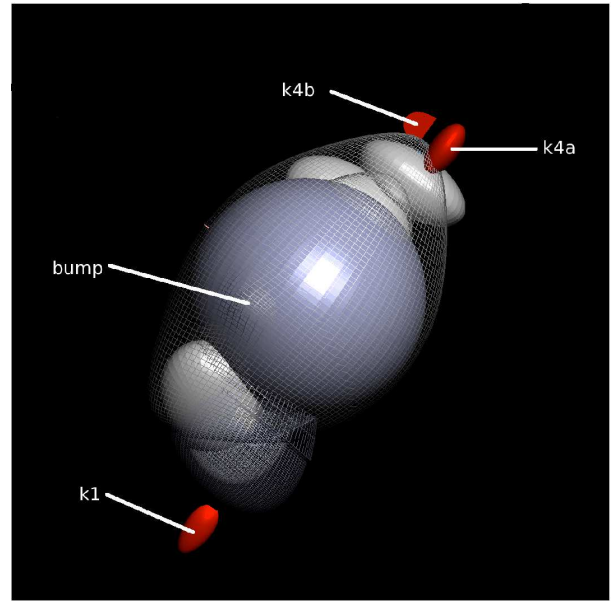}
	\caption{{\sc Shape} model of the inner structures. Knots and bump are highlighted.}
	\label{inner-model}
\end{figure}

\begin{figure}
 \centering
 \includegraphics[width=0.3\textwidth]{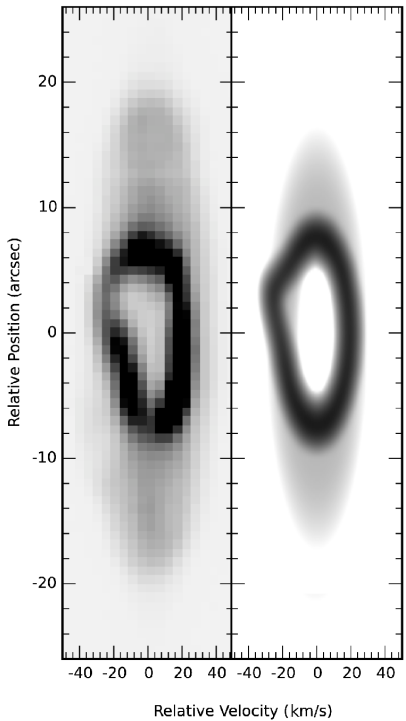}
\caption{Observed PV-map of the [O\,{\sc iii}] line at s4 and the corresponding {\sc shape} synthetic PV-map.
NE is up in both panels.}
\label{inner-shape}
\end{figure}


\begin{figure}
\centering
\includegraphics[width=0.45\textwidth]{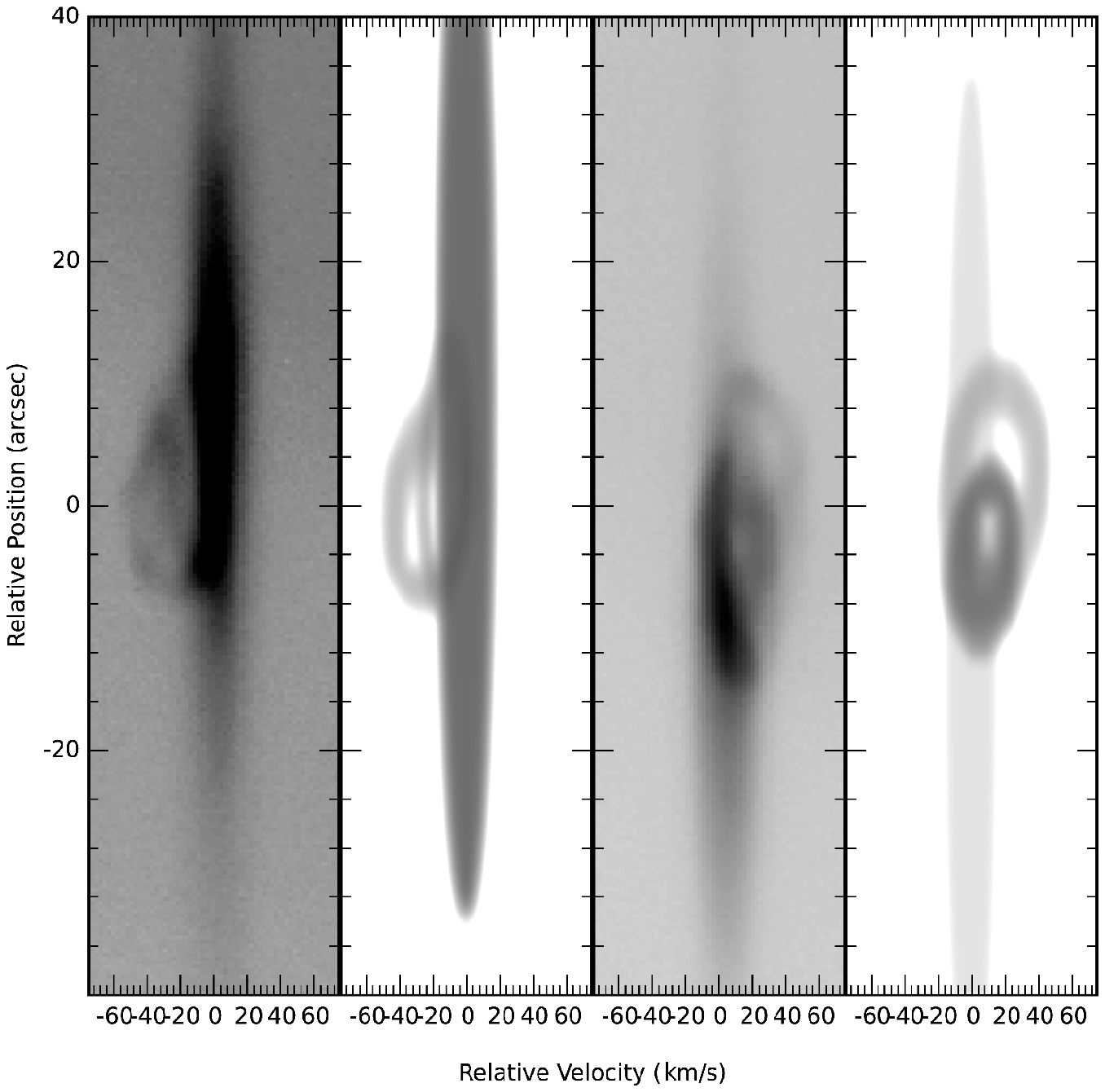}
\caption{Observed PV-map of the [O\,{\sc iii}] line for s2 and s12, and their corresponding {\sc shape} synthetic 
PV-maps. NE is up in all the panels.}
\label{bubbles-shape}
\end{figure}

We find that NGC\,3242 can be modeled as a double-shell elliptical
structure. Further, we have been able to calculate the
de-projected expansion velocities and radius for the inner shell as
well, resulting in values of $R_{\rm eq}=0.03\pm0.004$\,pc,
$V_{\rm eq}=23$\,km\,s$^{-1}$, $R_{\rm pol}=0.038\pm0.005$\,pc, and
$V_{\rm pol}=25.5$\,km\,s$^{-1}$, whereas the outer shell has an
expansion de-projected velocities and radius of  $R_{\rm eq}=0.06\pm0.008$\,pc,
$V_{\rm eq}=10$\,km\,s$^{-1}$, $R_{\rm pol}=0.074\pm0.01$\,pc, and
$V_{\rm pol}=13$\,km\,s$^{-1}$ (where eq=equatorial and pol=pole of the
ellipse). These values were calculated with observational data
presented in Sec.~\ref{results} and assuming homologous velocity
expansion ($V \propto r$). With the estimation of the distance using
the expansion parallax mentioned in section \ref{distance-parallax}, we
have been able to calculate the kinematical age of the shells
resulting in $1400\pm350$~yr and $5400\pm1300$~yr for the inner and
outer shells, respectively.

In Table~\ref{res-shape} we present properties, such as the
de-projected velocities, inclinations, and kinematical ages, derived
for the small-scale features described in the results and modeled using {\sc shape}.
In our model of NGC\,3242, we used an ellipsoid to generate the feature {\bf k1} presented in the
Figure \ref{jets-shape}. For {\bf k4}, we used a pair of ellipsoids in the model to
represent the {\bf k4a} and {\bf k4b} features detected in the s11 PV (see Figure \ref{inner-model}). The
velocities and the properties of {\bf k1} and {\bf k4a}, previously classified as
low ionisation structures \citep[LIS,][]{gon01}  and also fast, low ionisation emission regions
\citep[FLIERs,][]{bal98}, suggest a jet nature for
these two structures ($|\vec{v}|$=117\,km\,s$^{-1}$ and
$|\vec{v}|$=140\,km\,s$^{-1}$, respectively), whereas the elongated appearance observed and
modelled for {\bf k4b}, as well as its high de-projected velocity
($|\vec{v}|$=121\,km\,s$^{-1}$), point to the idea of this feature being originated
through a collision of the jet {\bf k4a} with the inner shell of NGC\,3242, or other disruptive mechanism
, but further detailed study will be needed to get insights about the nature of this structure.

It is important to make clear that we have no hard restrictions to choose the inclination angles
for the collimated structures. We made our best assumption given our data and model, as well as rational inferences,
but we are aware that other angles are possible.

On the other hand, a remarkable feature, seen as a deformation of the bright ellipse of the inner shell, 
located at $\simeq3$\arcsec from the middle, is clearly noted in the PV of s4. This feature cannot be easily identified in the images, but we can find its location by comparing PV maps with images. The corresponding
location appears as a less bright region in the high resolution image of HST (Fig. \ref{mosaico-shape}, right).
We have interpreted and modelled this feature as a kind of protrusion, blowing out from the nucleus of the
nebula, and deforming the velocity ellipse of the shell. We have used the modifier "bump" in {\sc Shape}
to model this protrusion with a Gaussian density profile to enhance it (Fig. \ref{inner-model}) to reproduce 
the deformation (Fig. \ref{inner-shape}). Possibly, this could be the youngest collimated 
ejection of the nebula, however the estimate of its kinematical age (1400~yr) is slightly lower than the age of the 
inner shell. This could be a matter of age and/or inclination angle uncertainties, but also could indicate a 
deceleration of this outflow when interacting with the inner shell.

To model the disruptions in the halo seen in Fig. \ref{umask2}, we used two pair of ellipsoids with a
Gaussian density profile for each bubble, reproducing well
these disruptions. The radial
velocities as well as the de-projected velocities calculated for these components
are remarkably high (see Table~\ref{res-shape}),
exceeding the 100\,km\,s$^{-1}$. A comparison between spectra and model of the disruptions
is shown in Fig. \ref{bubbles-shape}.

At first, the velocities and the spatial distribution of the polar
structures of NGC\,3242, the inner
knots, whose properties are in agreement with a FLIER nature, 
as well as the bubbles located between the double-shell
structure and the disruptions in the halo, suggest a common
origin. That is, the small-scale inner FLIERs and bubbles have compressed and pushed the
gas, reaching the halo and thus forming the disruptions observed on
these outermost regions. However, the kinematical ages estimated point
to separated events, being the inner structures younger than the ones
detected on the halo. In summary, the velocities and kinematical ages
of the FLIERs, bubbles, and disruptions suggest the presence of a mechanism that
triggers episodic ejections in this PN in association with rotation of
the collimating engine of NGC\,3242 \citep[e.g. NGC\,6543,][]{mir92}. 

\begin{table}
	\centering
\scalebox{0.9}{
	\begin{tabular}{|l|c|c|c|c|c|r|}
		\hline
		Name	&	$r$ 	& $V_{\rm rel}$ & $|\vec{V}|$ & PA &	$i$ 	&	$\tau_{\rm k}$\\ 
				&(arcsec)&(km\,s$^{-1}$)&(km\,s$^{-1}$)&($^{\circ}$)&($^{\circ}$)& (yr) \\ \hline
		k1		&	21	&	+22	&	117	&158	    &	100	&	580$\pm$120\\
		k4a$^*$	&	17	&	$-28$&	140	&$-43$ &	78 	&	390$\pm$100\\
		k4b$^*$	&	17	&	+13	&	121	&$-43$ &	96	&	450$\pm$100\\
		Bump	&	14	&	$-23$&	28	&124	    &	17	&	1600$\pm$400\\
		bh1		&	37	&	$-19$&	52	&131	    &	65	&	2280$\pm$450\\
		bh2		&	32	&	$-29$&	130	&143	    &75	&	800$\pm$200\\
		bh3		&	30	&	+29	&	140	&$-33$&102	&	690$\pm$150\\
		bh4		&	26	&	+13	&	83	&$-50$ &101	&	1000$\pm$250\\
		\hline
			\multicolumn{7}{|l|}{{\scriptsize $^*$k4a correspond to the blueshift knot and k4b to the 
			redshift knot in the spectra s11.}} \\
		\hline
	\end{tabular}
	
}      

	\caption{De-projected velocities, inclinations respect to the
          plane of the sky, and
          kinematical ages obtained from the main structures
          detected. The kinematical ages where determined using the
          distance calculated in the Section \ref{distance-parallax}
          as well as the de-projected velocities and radius obtained with the model. \label{res-shape}}
\end{table}
                     
\subsection{The formation of NGC\,3242}

We support the scenario in which the inner shell of NGC\,3242 was formed through the
shock front produced by the interaction of the stellar winds \citep{kwo78}, whereas the
outer shell can be interpreted as the AGB envelope remnant,
considering the kinematics hereby reported as well as the physical conditions reported in
previous studies \citep{mea00, ram09, nie11, mon13}. As for the kinematics of the inner FLIERs, bubbles, and
disruptions in the halo, we have
analysed them with unprecedented detail. The preference direction,
towards NW--SE (Fig. \ref{slits}), of
all these morphological components may be related with some mechanism
able to collimate such structures, likely in different events \citep[e.g.,][]{mir92,lop95,pal96,vaz08,rub15}. 

As stated before, from the morphological and kinematical data, we have
analysed for the first time the morpho-kinematic structure of the four disruptions in the
halo of NGC\,3242 ({\bf bh1}, {\bf bh2}, {\bf bh3}, {\bf bh4}). The small difference between the radial and the 
de-projected velocities of {\bf bh1} with respect to {\bf bh4}, as well as {\bf bh2} and {\bf bh3}, in addition to their 
kinematical ages suggest that these four structures were probably formed by the same phenomenon. This occurs 
when a collimated outflow passed through the outer shell breaking or pushing this shell as a result of the hight 
velocity of the jet. As a consequence, this leaves the diffuse disruptions as remnant of this collimated ejection
(differences could be originated by inhomogeneous medium). On 
the other hand, the inner FLIERs and bubbles were likely
formed by a later ejection, being the last morphological structures formed in
this PN.


In the case of the southern FLIER {\bf k1}, noticeably
orientated towards the south, not in the SE direction like the rest of
the structures, it is possible that {\bf k1} has followed this trajectory due a
deflection originated by higher density zones along the inner shell of
NGC\,3242. In this context, the value for the $N_e$ reported by \citet{nie06} for this shell is $\simeq$
2200\,cm$^{-3}$, however, the zones towards {\bf k1} and {\bf k4}, including the
knot {\bf k2}, have higher
electronic densities, up to 4000\,cm$^{-1}$ or greater, accordingly to the analysis
of the physical conditions carried out by \citet{mon13}. Besides the velocities hereby reported and the densities
measured in previous studies, the jet nature of the FLIERs {\bf k1} and {\bf k4} is favored considering
their emission in low-excitation emission lines, a well known
cooling mechanism of shocked gas. The knots {\bf k2} and {\bf k3}, emitting also in
high-excitation lines, may be produced by shocked material as well, although, the values for the electronic
temperature in these zones ($T_{\rm e}$=12,000 K) reported by \citet{mon13}
are not considerably higher than the values
obtained for the rest of the nebula (10,000\,K$>T_{\rm e}>$15,000\,K). On the other hand, there is clear 
evidence of filamentary structures
related with thermal dust and gas beyond the outer shell and the halo of this PN,
indicating that NGC\,3242 was born in a clumpy environment
\citep{ram09}, interacting with the surrounding ISM, likely affecting the
formation of some structures due to the presence of this dusty material.

\section{Conclusions}


 From the kinematic study we derived a systemic velocity for NGC\,3242 of 
$V_{\rm LSR}=-6.6\pm1$\,km\,s$^{-1}$. 
We also reassert that NGC\,3242 is a multiple-shell PN. The inner shell  has an expansion velocity of 
$\simeq$23\,km\,$s^{-1}$, meanwhile, the outer shell has an expansion velocity of $\simeq$13\,km\,s$^{-1}$. 
Further, beyond its multiple shell structure, we analysed the radial velocity of the 
disruptions in the halo located toward NW--SE along major axis of the PN, whereas the radial velocities are 29\,km\,s$^{-1}$ 
and 13\,km\,s$^{-1}$ and, $-19$\,km\,s$^{-1}$ and $-29$\,km\,s$^{-1}$,
respectively. Employing the expansion parallax method and using HST imaging from two different epochs, we have been able 
to estimated a new distance value of to the nebula of $d=660\pm100$\,pc. 


 A detailed morpho-kinematic model was built using the computational software
{\sc Shape} fitting our data and improving our interpretation of the structure of
NGC\,3242. Our model is consistent with an outer shell 
with a semi-major axis of 0.072\,pc ($V_{\rm pol}$=13\,km\,s$^{-1}$), 
possibly disrupted by collimated outflows 
passing through it, and an inner shell with semi-major axis of 0.037\,pc
($V_{\rm pol}$=25.5\,km\,s$^{-1}$), and likely
deformed by jets with a FLIER nature. According to our model, 
the disruptions in the halo have inclination angles of $i=65\degr$ and 
$i=75\degr$ toward SE, and $i=101\degr$ and $i=102\degr$ toward NW, where the de-projected velocities are 
52\,km\,s$^{-1}$, 130\,km\,s$^{-1}$, 83\,km\,s$^{-1}$, and 140\,km\,s$^{-1}$, respectively. Moreover, the jets in the 
inner shell have inclination angles of $i=100\degr$ and PA=+160\degr
and, $i=78\degr$ and PA=$-40$\degr, whereas 
the de-projected velocities are 117\,km\,s$^{-1}$ and 140\,km\,s$^{-1}$, respectively. The preferential direction, 
towards NW--SE of all these morphological components may be related with some mechanism able to collimate 
and eject such structures episodically. 

In our scenario, the outer shell is the oldest structure (5400~yr), then, some material was ejected and the first 
disruptions appear ({\bf bh1} and {\bf bh4}), protruding in the shell and partially in the halo. After that, the second 
disruptions were ejected (possibly at a higher velocity, {\bf bh2} and {\bf bh3}). The inner shell was ejected after the 
{\bf bh}s, and then the ``bump''. Finally, the features {\bf k} were
ejected and they interacted with the inner shell. All these results, as
well those derived in previous studies, suggest that all
the morphological components of NGC\,3242 apparently were originated in a clumpy medium.
Inconsistencies in ages of the inner shell and the
bump, as well as {\bf bh2}, {\bf bh3}, and {\bf bh4}, are small enough to be explain with age and inclination 
angle uncertainties.

\section*{Acknowledgments}

We are grateful to the staff of OAN-SPM, specially to Mr. Gustavo Melgoza-Kennedy, our telescope operator, for 
his assistantship during observations. This paper has been supported by Mexican grant IN107914 (PAPIIT-
DGAPA-UNAM). SZ acknowledges support from the UNAM-ITE collaboration agreement 1500-479-3-V-04. SA 
acknowledges support from the UNAM-UANL collaboration agreement. This research has made use of the 
SIMBAD database, operated at CDS, Strasbourg, France. IRAF is distributed by the National Optical Astronomy 
Observatories, which are operated by the Association of Universities for Research in Astronomy, Inc., under 
cooperative agreement with the National Science Foundation. MAGM acknowledges CONACYT for his graduate 
scholarship. Authors are deeply grateful to Prof. Michael Richer for a kind and careful revision of the manuscript.
Authors are grateful to the anonymous referee whose comments were truly valuable and helped to improve this
article.

\label{lastpage}

\end{document}